# A reduced cost equation of motion coupled cluster method for excited states based on state-specific natural orbitals: Theory, Implementation, Benchmark

Amrita Manna[a], Achintya Kumar Dutta[a,b]*

[a]*Department of Chemistry, Indian Institute of Technology Bombay, Powai, Mumbai 400076, India*

[b] *Department of Inorganic Chemistry, Faculty of Natural Sciences, Comenius University, Bratislava Ilkovičova 6, Mlynská dolina 842 15 Bratislava, Slovakia*

**Abstract:**

We present a reduced-cost equation-of-motion coupled-cluster method for excited states, built on a new state-specific frozen natural orbital (SS-FNO) framework. This approach enables systematic and controllable truncation of the virtual spaces, significantly reducing computational demands while maintaining high accuracy. The method allows black-box application via two adjustable thresholds and includes a perturbative correction that compensates for truncation errors. We have tested the performance of both CIS(D) and ADC(2) methods in generating appropriate natural orbitals for excited states. Benchmarking on valence, Rydberg, and charge-transfer excited states demonstrates excellent agreement with canonical EE-EOM-CCSD results, with mean absolute deviations typically below 0.02 eV when ADC(2) natural orbitals with perturbative corrections are applied.

**achintya@chem.iitb.ac.in*

## 1. Introduction:

Simulation of excited states is one of the most active branches of theoretical chemistry, which has implications in spectroscopy and photochemistry, as well as in emerging fields like energy conversion and storage. A variety of theoretical methods for the quantum chemical calculation of molecular excited states, starting from semi-empirical[1], time-dependent density functional theory(TDDFT)[2–5] to the state-of-the-art wave function-based methods[6], are available in the literature. Among all these methods, the TDDFT is most popular due to its low computational cost. However, it often fails for conjugated systems and charge-transfer states[5]. The wave function based methods are generally free from the above mention problems[6]. Among all the single reference wave function based methods described in the literature, the coupled-cluster method has emerged as one of the most accurate and systematically improvable one[7]. The ground state coupled-cluster is generally extended in excited states[8] by using the equation of motion(EOM) [9–12] approach. It is generally used in singles and doubles approximations[11] (EOM-CCSD), which scales as $N^6$ power of the basis function. The high computational scaling and the associated storage requirement of EOM-CCSD calculations prohibit its use beyond small molecules. A lot of effort has already been devoted towards reducing the computational cost of EOM-CC calculations[6]. All the lower scaling approximations to the EOM-CCSD can be broadly classified into three distinct categories. The first category of methods[13,14] involves approximating the one-electron and the two-electron integrals using Density-Fitting(DF)[15–21] approximations. This leads to a drastic reduction in storage requirements and introduces very small errors in standard calculations. However, the use of the standard density fitting method can reduce the computational scaling of only so-called 'Coulomb' kind of integrals and can not decrease the formal scaling of the coupled cluster method unless special techniques[22,23] are employed. The second category method employs perturbational truncation of the ground and excited state wave functions to reduce the computational cost of EOM-CCSD calculations[13,24,25]. However, as with other perturbative approximations, the accuracy of the methods depends upon the availability of a suitable zeroth-order reference state and does not provide uniform accuracy for all kinds of excited states[26]. The third kind of approximation involves the truncation of the wave function using localized and natural orbitals. One of the first lower scaling approximations to the EOM-CCSD method, based on natural transition orbitals, is reported by Mata and Stoll[27]. Korona and co-workers[28] as well as Crawford and King[29] have reported localized orbital based implementation of EOM-CCSD. Hattig and co-workers have reported state-specific pair natural orbital based CC2[30], EOM-CCSD[31], and EOM-CC3[32] methods, which can be applied beyond small molecules. Neese and co-workers have proposed an alternative approach to reduce the computational cost of ground-state coupled-cluster methods by employing pair natural orbitals[33]. Additionally, they utilized the technique of successive similarity transformation[34,35] to decrease the computational cost of excited-state calculations. A lower scaling coupled cluster based excited state method based on state-specific natural orbitals has been reported by Kallay and co-workers[36] and Koch and co-workers[37]. Kallay and co-workers have extended the natural orbital-based lower scaling approximation to Algebraic Diagrammatic Construction (ADC) methods.[38–40] Valeev and co-workers[41], on the other hand, used state-averaged natural orbitals. Among the various flavors of natural orbitals available in the literature, the Frozen Natural Orbitals[42] (FNO) is one of the most popular options. However, most of the existing FNO-based implementation[43,44] of EOM-CC methods use a single set of natural orbitals calculated from the MP2 ground state method. However, to extend to excited states, one needs to develop a state-specific FNO framework. The aim of this manuscript is to describe the theory, implementation, and benchmarking of a reduced cost EOM-CCSD method based on state-specific FNOs.

## 2. Theory:

## 2.A. Equation of motion coupled cluster (EOM-CC) theory:

We will represent the following convention for classifying different types of indices.

● i, j, k, l: Indexes of occupied molecular orbitals in Hartree-Fock determinant.

● a, b, c, d: Indexes of unoccupied (virtual) molecular orbitals.

● p, q, r, s: General indexes of any molecular orbital.

● P, Q: Indexes density fitting auxiliary functions.

Schrödinger equation for the single-reference coupled-cluster equation can be expressed as

$$\hat{H}e^{\hat{T}}\left|\phi_0\right\rangle = E_{cc}e^{\hat{T}}\left|\phi_0\right\rangle \quad (1)$$

Where $\phi_0$ is the zeroth-order wave function, which is generally but not necessarily a Hartree-Fock determinant. The $\hat{T}$ is the cluster operator, which can be written as

$$\hat{T} = \hat{T}_1 + \hat{T}_2 + \hat{T}_3 + ....... + \hat{T}_N \quad (2)$$

With $\hat{T}_m = \frac{1}{(m!)^2} \sum_{\substack{ij... \\ ab...}} t_{ij....}^{ab...} \hat{a}_a^\dagger \hat{a}_b^\dagger ......\hat{a}_j \hat{a}_i$ ; where $t_{ij...}^{ab...}$ are the cluster amplitudes and $\hat{a}^\dagger$, $\hat{a}$ are creation and annihilation operators in the second quantization notation.

The $\hat{T}$ operators are generally used in single and double approximations (CCSD),

$$\begin{aligned} \hat{T} &= \hat{T}_1 + \hat{T}_2 \\ &= \sum_{i,a} t_i^a \left\{ \hat{a}_a^\dagger \hat{a}_i \right\} + \frac{1}{4} \sum_{i,j,a,b} t_{ij}^{ab} \left\{ \hat{a}_a^\dagger \hat{a}_i \hat{a}_b^\dagger \hat{a}_j \right\} \end{aligned} \quad (3)$$

After left multiplication by $e^{-\hat{T}}$ in equation (1), one can arrive at

$$\bar{H} \left| \phi_0 \right\rangle = E_{cc} \left| \phi_0 \right\rangle \, (4)$$

Where $\bar{H} = e^{-\hat{T}} H e^{\hat{T}}$ is the coupled cluster similarity transformed Hamiltonian and $E_{cc}$ is the ground state coupled cluster energy. The solutions for $E_{cc}$ and $t$ amplitudes are obtained by iterative solutions of the following sets of equations,

$$\left\langle \phi_0 \,|\, e^{-\hat{T}} H e^{\hat{T}} \,|\, \phi_0 \right\rangle = E_{cc} \, (5)$$

$$\left\langle \phi_{ij...}^{ab...} \,|\, e^{-\hat{T}} H e^{\hat{T}} \,|\, \phi_0 \right\rangle = 0 \, (6)$$

The coupled cluster method is generally extended to excited states using the EOM formalism, which uses a CI like linear excited operator $\hat{R}$

$$\bar{H} \hat{R}_k \left| \phi_{cc} \right\rangle = E_k \hat{R}_k \left| \phi_{cc} \right\rangle \, (7)$$

$E_k$ is the energy of the k$^{th}$ target state and $\hat{R}_k$ is the corresponding excitation operator. One can directly compute the difference of excitation energy $\omega_k = E_k - E_0$ from the commutator form of the equation (7)

$$\left[ \bar{H}, \hat{R}_k \right] \left| \phi_0 \right\rangle = \omega_k \hat{R}_k \left| \phi_0 \right\rangle \, (8)$$

The form of $\hat{R}_k$ depends upon the nature of the target state, and for excited states it has the following form

$$\hat{R}_k^{EE} = r_0 + \sum_{i,a} r_i^a \left\{ \hat{a}_a^\dagger \hat{a}_i \right\} + \sum_{i<j,a<b} r_{ij}^{ab} \left\{ \hat{a}_a^\dagger \hat{a}_i \hat{a}_b^\dagger \hat{a}_j \right\} \quad (9)$$

The equation (8) is solved by the modified Davidson iterative diagonalization method[45], which involves the contraction of the similarity transformed Hamiltonian with suitably chosen guess vectors to generate the so-called sigma vectors. Here $\bar{H}$ is non-Hermitian, so one can obtain a bi-orthonormal set of right and left eigen vectors[8]. However, it is sufficient to calculate only one of them to obtain the excitation energy. The EOM part for excitation energy scales as the $N^6$ power of the basis set for the EE problem, similar to the ground state coupled cluster method. Routinely performing EOM-CCSD calculations beyond molecules containing more than ten non-hydrogen atoms and lacking point group symmetry can pose challenges unless additional approximations are made or access to a high-performance supercomputing facility is available.

## 2.B. Frozen Natural Orbitals (FNOs):

Natural orbitals (NOs) are formed from one body-reduced correlated density matrix[46]. Among the various flavors of natural orbitals available, we have used the frozen natural orbital[42] (FNO) approximations where the occupied orbitals are kept frozen at their HF descriptions, and virtual orbitals are expanded in terms of natural orbitals. The MP2 method[47] is generally used to obtain the one-body reduced density matrix for FNOs. The virtual-virtual block of the one-body reduced density matrix is constructed using the MP2 method as

$$D_{ab}^{MP2} = \sum_{ijc} \left( 2T_{ij}^{ca} T_{ij}^{cb} - T_{ij}^{ca} T_{ij}^{bc} \right) (10)$$

Where $T_{ij}^{ab}$ is a MP2 doubles amplitude defined as

$$T_{ij}^{ab} = \frac{(ai|bj)}{\varepsilon_i + \varepsilon_j - \varepsilon_a - \varepsilon_b} (11)$$

Where $\varepsilon$ are Hartree-Fock orbital energy. By diagonalizing this density matrix obtained from equation 10, one can get the FNOs as $V$ and the corresponding occupation numbers denoted by $n$

$$D_{ab}^{MP2} V = Vn \; (12)$$

The FNOs with lower occupation numbers make a negligible contribution towards correlation energy, and the corresponding virtual orbitals can be dropped from the correlation calculation.

Consequently, one can choose a predefined threshold, the natural orbitals below which can be removed from the virtual space. The virtual-virtual block of the Fock matrix is then transformed into a truncated natural orbital basis,

$$\tilde{F}_{vv} = \tilde{V}^{\dagger} F_{vv} \tilde{V} \ (13)$$

Where $F_{vv}$ the virtual-virtual block of the initial matrix, $\tilde{V}$ is the truncated natural virtual orbital. By diagonalizing this semi-canonical Fock matrix, we get orbital energy $\tilde{\varepsilon}$ in a truncated virtual natural orbital basis,

$$\tilde{F}_{vv} \tilde{Z} = \tilde{Z} \tilde{\varepsilon} \ (14)$$

The matrices

$$\tilde{U}_{vir} = U_{vir} \tilde{V} \tilde{Z} \ (15)$$

$$\tilde{U}_{occ} = U_{occ} \ (16)$$

denotes the transformation from canonical space to semi-canonical Frozen natural orbital space, respectively.

## 2.C. State-Specific Frozen Natural orbitals (SS-FNOs):

In most of the FNO-based implementations of CC and EOM methods, a single set of FNOs generated from a ground-state MP2 calculation is employed. However, the electron distributions in the excited state may be significantly different from the ground state. Consequently, using separate sets of natural orbitals for ground and excited states is justified. The use of state-specific natural orbitals has been advocated by Hattig and co-workers[48] as well as Kallay and co-workers[36]. Valeev and co-workers[41], on the other hand, used a state-averaged natural orbital, which can be advantageous for calculations of multiple roots in a single excitation. Nevertheless, employing state-averaged natural orbitals may result in a large size of the virtual space when the characteristics of concurrently calculated excited states vary considerably. Both state-specific and state-averaged natural orbitals described above are calculated using the CIS(D) approximations. One can alternatively obtain an approximation to the EOM-CCSD excitation energies, which is complete up to 2$^{nd}$ order in perturbation by diagonalizing the approximate CC jacobian

$$A^{(2)} = \begin{pmatrix} \left\langle \mu_1 \,|\left[ \left( \hat{H} + \left[ \hat{H}, T_2^{(1)} \right] \right), \tau_{\nu_1} \right] |\, HF \right\rangle & \left\langle \mu_1 \,|\left[ \hat{H}, \tau_{\nu_2} \right] |\, HF \right\rangle \\ \left\langle \mu_2 \,|\left[ \hat{H}, \tau_{\nu_1} \right] |\, HF \right\rangle & \left\langle \mu_2 \,|\left[ \hat{F}, \tau_{\nu_2} \right] |\, HF \right\rangle \end{pmatrix} \quad (17)$$

The obtained energies will be identical to CIS(D $\infty$) method of Head Gordon and co-workers[49]. However, the form of $A^{(2)}$ as described in equation (17) is non-Hermitian, and one needs to calculate both right and left eigenvectors to obtain the specific one-particle reduced density matrix. A Hermitian approximation to $A^{(2)}$ can be obtained as

$$A_{ADC(2)}^{(2)} = \frac{1}{2}\left( A^{(2)} + \left( A^{(2)} \right)^{\dagger} \right) \quad (18)$$

which is identical to the second-order algebraic diagrammatic construction (ADC(2)) method[50]. In this work, we have used the ADC(2) method instead of CIS(D $\infty$) due to the ease of calculation of density in the former. It should be noted that one can obtain the CIS(D) excitation energy as a non-iterative approximation to equation (17)

$$\omega_{CIS(D)} = \omega_{CIS} + R_{CIS}^{\dagger} A_{SD} \left( \omega_{CIS} I - F \right)^{-1} A_{DS} R_{CIS} \quad (19)$$

Where $\omega_{CIS}$ and $R_{CIS}$ are the CIS eigen values and eigen vectors, respectively, and $F$ is the Fock matrix. The ADC(2) natural orbitals are expected to provide higher accuracy than the CIS(D) natural orbitals due to a more complete treatment of the excited state wave function in the former.

In the state-specific FNO(SS-FNO) approach, two separate sets of frozen natural orbitals are used for each excited state calculation, one for the reference state wave-function and another for the target state wave-function. The target state frozen natural orbitals are generated by diagonalizing the virtual-virtual block of the second-order approximate one-particle reduced density matrix of the target state calculated from the CIS(D) or ADC(2) method.

$$D_{ab}^{EE} = D_{ab}^{MP2} + D_{ab}^{ADC(2)/CIS(D)} \quad (20)$$

The most time-consuming part in the construction of $D_{ab}^{MP2}$ is the assembly of the integral list $(ai\,|\,bj)$, which scales with a fifth-power complexity $n_{occ}^2 n_{vir}^2 n_{aux}$, where $n_{aux}$ is the dimension of auxiliary basis function used in the density fitting approximation. Another major computational expense is the evaluation of the virtual-virtual block of the density matrix, which also has a fifth-power scaling $n_{occ}^2 n_{vir}^3$. It is worth noting that the MP2 densities only need to be constructed once, regardless of the number of excited states. The excited state density matrix can be decomposed into two parts: the CIS density matrix and the 2nd order correction

$$D^{ADC(2)/CIS(D)} = D^{CIS} + D^{(2)} \quad (21)$$

The CIS density matrix is expressed as a sum of CIS coefficients

$$D_{ab}^{CIS} = \sum_i c_i^a c_i^b \quad (22)$$

Where $c_i^a$ is a CIS coefficient for a particular state, while the doubles contribution to the density matrix can be calculated in a similar manner to the MP2 counterpart

$$D_{ab}^{(2)} = \sum_{ijc} \left( 2c_{ij}^{ca} c_{ij}^{cb} - c_{ij}^{ca} c_{ij}^{bc} \right) \quad (23)$$

Where $c_{ij}^{ab}$ is the ADC(2) or CIS(D) doubles amplitude. In the language of the ADC method, equation (22) and (23 ) will be equivalent to calculating excited state density using zeroth-order intermediate state representation with ADC(2) vectors[40]. The most time-consuming step in the excited state density matrix is the solution of ADC(2) equation, the most expensive term of which involves contraction of $c_{ij}^{ab}$ with the three external integrals and it scales as iterative $n_{occ}^2 n_{vir}^2 n_{aux}$ (non iterative $n_{occ}^2 n_{vir}^2 n_{aux}$ of CIS(D)).

The construction of the virtual -virtual block of the density matrix for excited state scales as $n_{occ}^2 n_{vir}^3$. The excited state FNOs are obtained from $D_{ab}^{EE}$ follow equation 11 to 16. The transformation matrix

$$Y = \tilde{U}_{GR}^{\dagger} \tilde{U}_{EE} \quad (24)$$

connects the ground and excited state FNO basis.

## 2.D. Density Fitting (DF) and Natural Auxiliary Functions (NAFs):

Introducing the DF approximation[36], the four-centered two-electron repulsion integrals (ERIs) can be written in the

$$(pq|rs) = \sum_{PQ} (pq|P)(P|Q)^{-1}(Q|rs) \quad (25)$$

form, where $P$ and $Q$ stand for the elements of the auxiliary basis, whereas $(pq|P)$ and $(P|Q)$ are the three- and two-center Coulomb integrals, respectively, and $(P|Q)^{-1}$ is a simplified notation for the corresponding element of the inverse of the two-center Coulomb integral matrix. In practice, the latter inverse matrix is factorized and rewritten, e.g., as the product of the inverse square root matrices. The four-center ERIs can then be recast in the form

$$(pq|rs) = \sum_{Q} J_{pq}^{Q} J_{rs}^{Q} \quad (26)$$

where

$$J_{pq}^{Q} = \sum_{P} (pq|P)(P|Q)^{-1/2} \quad (27)$$

All the integrals in the present implementation are generated using the density fitting approximation. The integrals of up to two virtual indices are constructed and stored in memory/disk. The integrals containing three and four virtual indices are not stored, and the terms that involve the contraction of $t$ or $r$ intermediates with integrals with three or four virtual indices are factorized in terms of the three-centered $J_{pq}^{Q}$.

As the natural orbitals are used to truncate and reduce the size of the orbital basis set, one can use the natural auxiliary function (NAF) approach of Kallay and co-workers[51] to reduce the dimension of the auxiliary basis set. In the NAF approach, a singular value decomposition of the three-center two-electron matrix is performed.

$$J = M \Sigma N^{T} \quad (28)$$

Where $M$ and $N$ are the matrices composed of the left singular vectors, and $\Sigma$ is a diagonal matrix made up of the singular values in its diagonal. The right-side SVs of $J$ can be considered as the linear combination of the coefficients of the natural auxiliary functions. The $N$ and $\Sigma$ are identical to the square root of the eigenvalues and eigenvectors of the matrix

$$W = JJ^{T} \quad (29)$$

and can be more efficiently obtained by diagonalizing $W$. The NAFs with eigenvalues lower than a threshold ($\varepsilon_{NAF}$) are dropped, and the remaining $\bar{n}_{aux}$ eigenvectors are collected in the matrix $\bar{N}$. The auxiliary function index of matrix $J$ is now transformed by matrix $\bar{N}$ as

$$\bar{J} = J\bar{N} \quad (30)$$

The resulting $\bar{J}$ matrix is the best rank $\bar{n}_{aux}$ representation of $J$ in the least-squares sense[51]. The most computation-intensive operation of SS-FNO-EE-EOM-CCSD calculation scales as $n_{occ}^2 n_{vir}^3 n_{aux}$, which involves contraction of the four virtual integrals with $t_2$ or $r$ amplitudes. Consequently, the decrease in the computation time in the NAF basis is expected to be proportional to the ratio $\frac{n_{aux}}{\bar{n}_{aux}}$. One can use any of the three-center two-electron integrals $J_{ia}^{Q}$ and $J_{ab}^{Q}$ for the construction of natural auxiliary functions. Following the suggestion of Kallay and co-workers[36] we have used the $J_{ab}^{Q}$ integrals for generating natural auxiliary functions.

## 2.E. Perturbative correction:

Perturbative corrections for the natural orbital truncation have been shown to significantly improve the accuracy of natural orbital based ground state coupled cluster calculations[47,52,53]. Similarly, one can perform a perturbative correction for the truncated virtual space in the SS-FNO-EE-EOM-CCSD calculation. However, the effect of perturbative truncation has not been considered for almost all the natural orbitals-based implementations of the excited state coupled cluster method[30–33,36,41]. We employed the ADC(2) method, used to generate SS-FNO, to apply a perturbative correction for the truncated virtual space and the truncated natural auxiliary function. The corrected EE value in SS-FNO-EE-EOM-CCSD is defined as

$$\omega_{EE-EOM-CCSD} \approx \omega_{SS-FNO-EE-EOM-CCSD}^{corrected} = \omega_{SS-FNO-EE-EOM-CCSD}^{uncorrected} + \left(\omega_{EE-ADC(2)}^{canonical} - \omega_{EE-ADC(2)}^{SS-FNO}\right) \quad (31)$$

## 2.F. General Algorithm:

There is an overview of our general algorithm for the SS-FNO-EE-EOM-CCSD method, which is as follows:

1. Solve the HF equations.
2. Generate the three-centre two-electron integrals using DF approximations.
3. Solve the MP2, CIS and EE-ADC(2)/CIS(D) equations for all the excited state in canonical basis. Save the amplitudes to disk.

4. Generate the ground state frozen natural orbitals by the MP2 density matrix and transform the molecular orbital space to frozen natural orbital basis by using the transformation matrix $\tilde{U}_{GR}$.

   4a. Transform the virtual MO indices $J$ to the ground state FNO basis $(\tilde{J})$.

   4b. Generate the $W$ from $\tilde{J}$, and transform the auxiliary index of $\tilde{J}$ to NAF basis.

   4c. Solve the ground state CCSD equations and save the amplitudes to disk.

5. Loop over excited states obtained from EE-ADC(2)/CIS(D) method

   5a. Form the excited state density matrix from equation (20).

   5b. Generate the excited state FNO and molecular orbital to the excited state frozen natural orbital transformation matrix $(\tilde{U}_{EE})$.

   5c. Transform the virtual MO indices of $J$ to the excited state FNO basis $(\tilde{J})$.

   5d. Generate the $W$ from $\tilde{J}$, and transform the auxiliary index of $\tilde{J}$ to NAF basis.

   5e. Calculate $Y$, retrieve the CCSD amplitudes from the disk, and transform them to excited state FNO basis using $Y$.

   5f. Calculate $\bar{H}$ and solve the EE-EOM-CCSD equations in the excited state FNO basis.

6. Retrieve the canonical MP2 amplitudes from the disk, transform them to the excited state FNO basis, solve EE-ADC(2)/CIS(D) equations, and calculate the perturbative correction to SS-FNO-EE-EOM-CCSD values.

In the EE-EOM-CCSD method, the total computational cost is dominated by both ground state CCSD and excited state EOM calculation, in which the most time-consuming step scales as iterative $(n_{aux})(nv)^3\left(no\right)^2$ where $n_{aux}$ is the number of auxiliary basis function and $nv$ is the number of virtual orbitals. The use of FNO and natural auxiliary function reduced the computational cost of the most time-consuming step to $(n_{aux}-n_{aux}^{f})(nv-nv_{f})^3\left(no\right)^2$ where $n_{aux}^{f}$ denotes the number of frozen auxiliary basis and $nv_{f}$ is the number of discarded virtual orbitals.

The calculations of FNO involve additional $(nv)^3(no)^2$ scaling steps, which generally negligible overhead in terms of total computational time.

The SS-FNO-EE-EOM-CCSD as described above is implemented in our inhouse software package BAGH[54]. BAGH is written primarily in Python with the computational bottleneck steps written in Fortran and Cython, and it can perform wave-function-based calculations based on Schrödinger and Dirac equations. BAGH relies on other software packages for the generation of one- and two-electron integrals and is currently interfaced with PySCF[55–57], GAMESS-US[58], and DIRAC[59]. The results presented in this paper are calculated using the PySCF interface of BAGH.

## 3. Result & Discussion:

## 3.1. Convergence with respect to the threshold:

The excited states of atoms and molecules can be broadly classified into three categories: valence, Rydberg, and charge transfer. In the ideal case, any approximations to the EE-EOM-CCSD method should give similar accuracy for all kinds of excited states. The accuracy of the SS-FNO-EE-EOM-CCSD method depends on the FNO truncation threshold of both ground and excited states. One needs to benchmark separately to arrive at an optimal FNO threshold for both ground and excited states. Figure 1 presents the convergence of the excitation energy to the corresponding $1^1B_2$ state of the pyrrole with respect to the size of the virtual space in different variants of the SS-FNO and canonical basis. The ground state FNO threshold and the natural auxiliary threshold have been set to zero. An aug-cc-pVTZ basis set was used for the calculations, and the canonical EE-EOM-CCSD result was used as a reference. The geometry of the pyrrole has been taken from ref[60]. It can be seen that the convergence of excitation energy is extremely slow with respect to the size of the virtual space on the canonical basis, and the results converge only on reaching 100% of the virtual space. The convergence is faster on the SS-FNO basis. The error comes down below 0.1 eV at around 50% of the virtual space, which is less than the error bar EOM-CCSD method. Both CIS(D) and ADC(2) give a similar performance and result in convergence at around 60% of the virtual space. However, at a lower size of the virtual space, the ADC(2) gives slightly better performance.

The FNO threshold is a more systematic criterion than the percentage of the virtual space. Figure 2(a) presents the convergence of error for the SS-FNO-EE-EOM-CCSD method for $1^1B_2$ excited state of pyrrole with respect to the excited state FNO threshold in the aug-cc-pVTZ basis set. The ground state threshold has been kept at zero. The canonical EE-EOM-CCSD result with 100% active virtual orbitals has been used as the reference. It can be seen that the excitation energy in the SS-FNO method converges systematically with respect to the excited state natural truncation threshold when ADC(2) natural orbitals are used and approaches the canonical result at $10^{-5}$ threshold. The inclusion of perturbative correction in SS-FNO-EE-EOM-CCSD shows very significant improvement over the uncorrected version and the results converge to canonical values at $10^{-4}$ threshold. The convergence of the excitation energy is a little slower when CIS(D) natural orbitals are used. However, the excitation energy in CIS(D) natural orbital based SS-FNO-EE-EOM-CCSD method also approaches the canonical result at the $10^{-5}$ threshold. At the $10^{-5}$ threshold, our FNO truncation scheme (See Table S1) selects 79% and 77% of the virtual space for ADC(2) and CIS(D) natural orbitals, respectively. At the $10^{-4}$ threshold, the use of ADC(2) excited state natural orbital selects only 47% of the virtual, which displays the computational advantage gained by the use of perturbative correction. Figure 2(b) presents the convergence of error for the SS-FNO-EE-EOM-CCSD method with respect to the ground state FNO threshold of the $1^1B^2$ state of the pyrrole molecule in the aug-cc-pVTZ basis set. Excited state FNO threshold is fixed at $10^{-4}$ threshold. The canonical EE-EOM-CCSD result with 100% active virtual orbitals has been used as the reference. It can be seen that both ADC(2) and CIS(D) based SS-FNO-EOM-CCSD method shows similar convergence behaviour with respect to ground state FNO threshold, and the results in both methods converge at $10^{-4}$ threshold. It selects 43% of the virtual orbital in both methods. The inclusion of perturbative correction improves the convergence behaviour and results in the ADC(2) method converging at $10^{-3}$ ground state FNO threshold, which selects only 17% of the virtual space. We are going to use the $10^{-4}$ threshold for both ground and excited states for all subsequent calculations.

To determine the optimal natural auxiliary function (NAF) threshold, we plotted the excitation energy of $1^1B_2$ state of pyrrole in SS-FNO-EE-EOM-CCSD at different NAF thresholds. Throughout these calculations, the ground and excited FNO threshold is fixed at $10^{-4}$ (see Figure S1). The aug-cc-pVTZ basis set has been used for the calculations. The auxiliary basis set is automatically generated from the PYSCF & gives 845 auxiliary basis functions for pyrrole. Our result shows that using a NAF truncation threshold of $10^{-2}$ leads to an error of less than 0.01 eV and retains only 48% of the auxiliary basis functions. This suggests that the use of NAFs can effectively reduce the computational cost of density-fitting-based EOM calculations. The perturbative correction seems to have a negligible effect on the accuracy for the truncation in NAF. Therefore, for the rest of the manuscript, we used the default value of $10^{-2}$ for the NAF threshold for the excited and ground states.

## 3.2: Benchmarking on extended test set:

### 3.2.1: Valence Excited States:

The accuracy of the excited states has been extensively benchmarked for Valence excited states due to the availability of a well-designed benchmark set by Thiel and colleagues[60]. A statistical analysis of SS-FNO-EE-EOM-CCSD results is presented in Table 1. TZVP basis set has been used for the calculations. The untruncated canonical EOM-CCSD results have been used as the reference. The EE values for all the molecules have been provided (supporting information). The maximum absolute deviation (MAD) observed in the SS-FNO-EE-EOM-CCSD result with CIS(D) SS-FNO is 0.045 eV, and it improves by 0.030 eV when the ADC(2) SS-FNO is used. The addition of the perturbative correction improves the result, and the MAD value reduces to 0.020 eV. The RMSD error in the SS-FNO-EE-EOM-CCSD method with CIS(D) SS-FNO is 0.009 eV and is reduced to 0.006 eV with ADC(2) SS-FNO. The error further reduces to 0.004 eV with perturbative correction. This trend is similar for other statistical parameters. The error distribution plot in Figure 3 shows the truncation of virtual space in the FNO basis leads to overestimation of excitation energy relative to the untruncated values. Both the magnitude and spread of the error decrease on using ADC(2) natural orbitals.

### 3.2.2: Charge-Transfer Excited States:

It is well-known that TD-DFT and many approximate excited-state methods struggle with charge transfer states. Hence, it is of interest to assess the performance of the SS-FNO-EE-EOM-CCSD method for charge transfer excited states. To explore the impact of charge transfer separability on the accuracy of excitation energy for these states, the benchmark set proposed by Kozma and co-workers[61] was used. This set consists of ten dimers with low-energy CT states, and the corresponding excitation energies were calculated using the cc-pVDZ basis set. The reference EOM-CCSD results for all 14 CT states are from Ref. 61, and the statistical analysis of these results is presented in Table 2. The MAE and MAD (maximum absolute deviation) of SS-FNO-EE-EOM-CCSD using ADC(2) SS-FNO are 0.018 eV and 0.083 eV, respectively, which are almost identical to the value using CIS(D) density. These errors decrease with perturbative correction in SS-FNO-EE-EOM-CCSD results, which exhibit MAE and MAD values of 0.008 eV and 0.038 eV, respectively. The error distribution plot in Figure 4 demonstrates that charge transfer excitation energies are underestimated using ADC(2) and CIS(D) SS-FNO, which is in contrast with that observed for the valence excited states. The use of ADC(2) and CIS(D) natural orbitals results in a very similar distribution of errors. The inclusion of perturbative corrections leads to an improvement in both the magnitude and the spread of error.

### 3.2.3: Rydberg Excited States:

In order to evaluate the performance of the SS-FNO-EE-EOM-CCSD method for singlet Rydberg excited states, the Waterloo test set of Nooijen[62] was employed. The calculations were performed using Ahlrich's TZVP basis set, supplemented with additional 5s4p4d diffuse functions added to the ghost atom placed at the center of symmetry. The untruncated canonical EE-EOM-CCSD results serve as the reference values for the singlet states, and the statistical analysis of these results is presented in Table 3. From Figure 5, it can be seen that the use of the truncated SS-FNO basis overestimates the excitation energy for both CIS(D) and ADC(2) natural orbitals. The MAD value of the SS-FNO-EE-EOM-CCSD result is 0.121 eV using CIS(D) density. The value of MAD decreases to 0.069 eV when using ADC(2) density. The RMSD and MAE values are 0.021 and 0.014 eV using CIS(D) density. The error reduces to 0.017 and 0.013 eV, respectively, when ADC(2) SS-FNO is used. This trend is similar for other statistical parameters. The use of perturbative corrections leads to a slight improvement in the magnitude and spread of the error.

## 3.3. Computational efficiency and application on a large molecule:

To assess the efficiency of our implementation of the SS-FNO-EE-EOM-CCSD method, we performed calculations on a series of water clusters ($(H2O)_n$, n = 1-5) using the aug-cc-pVTZ basis set. The lowest excited state has been calculated for each cluster size. The calculations were carried out in both the SS-FNO and canonical EE-EOM-CCSD methods. Figure 6(a) presents the computational time required for these calculations vs the cluster size. The calculations were performed on a dedicated node with two Intel(R) Xeon(R) CPU E5-2620 v4 @ 2.10GHz and 94 GB of RAM. It can be seen that the computational timing for the SS-FNO-EE-EOM-CCSD method increases less steeply compared to the canonical version as the cluster size grows. In Figure 6(b), we have analysed the time required for various steps in the correlation calculation within the SS-FNO based implementation of EE-EOM-CCSD for the same series of water clusters at $10^{-4}$ threshold. It can be seen that the generation of ground and excited state FNO takes a negligible fragment of the total computational time, which is consistent with our analysis presented in section 2.F. The most time-consuming part is the ground state CCSD and excited state EOM-CCSD calculation.

To further investigate the performance of the SS-FNO-EE-EOM-CCSD method, we have calculated the first two excitation energies of a medium-sized molecule: N-methyl-2,3-benzcarbaxole. The molecule consists of 31 atoms and 122 electrons, and its geometry was obtained from the ref 36. We have used the aug-cc-pVTZ basis set and the aug-cc-pVDZ-ri auxiliary basis set, which resulted in a virtual space containing 1066 orbitals and an auxiliary basis set with dimensions of 1595. We have used a FNO truncation threshold of $10^{-4}$ for both the ground and excited states. It leads to a truncated virtual space of 449 for the ground state, and 457 and 458 for the two excited states. It gives an excitation energy of 3.63 eV and 4.23 eV, respectively. Our calculations were performed on a dedicated workstation equipped with 2 Intel Gold 5315 CPUs, providing a total of 16 cores and 2 TB of RAM. The entire calculation took 2 days, 16 hours to complete, of which 28 minutes were required for the Hartree-Fock calculations. The canonical EE-ADC(2) calculation for the first two excited states took 12 hours, 4 minutes. The CCSD calculation in the ground state FNO basis required 19 hours, 38 minutes. The EE-EOM-CCSD calculation in the SS-FNO basis for two roots took 31 hours and 35 minutes. The dominant part of the time in both ground and excited state calculations is spent on the contraction of the amplitude with the four external integrals. The treatment of those terms via tensor hypercontraction reduces the total time to 1 day 13 hours, out of which the ground state coupled cluster took 5 hours, and 17 hours 29 minutes were spent in the EOM step. More details about the THC-based implementation of four external integrals in the SS-FNO-EOM-CCSD method can be found in the supporting information. The error introduced by the THC is negligible, and the excitation energies obtained for the two states are 3.63 eV and 4.22 eV, respectively.

## 4: Conclusions:

We have developed and implemented a reduced-cost equation-of-motion coupled-cluster method for excited states by employing state-specific frozen natural orbitals (SS-FNOs) and natural auxiliary functions (NAFs). This approach enables a systematic and controllable truncation of the virtual orbital and auxiliary function spaces while maintaining high accuracy in excitation energy predictions. We have tested the performance of both CIS(D) and ADC(2) methods for generating state-specific natural orbitals and included a perturbative correction to compensate for the orbital and auxiliary space truncation.

Extensive benchmarks on valence, charge-transfer, and Rydberg excitation energy test sets demonstrate that the SS-FNO-EE-EOM-CCSD method achieves excellent agreement with canonical EE-EOM-CCSD results. The perturbative correction significantly enhances accuracy, especially at moderate truncation thresholds, without introducing cost. Moreover, the method exhibits nearly black-box performance with consistent behaviour across diverse excitation types.

In addition to achieving high accuracy, the SS-FNO-EE-EOM-CCSD method offers substantial computational savings. Timing analyses on water clusters and a medium-sized organic molecule (N-methyl-2,3-benzcarbazole) show that the method can be used for systems where canonical calculations are prohibitive. The SS-FNO scheme is general and can be used to reduce the computation cost of other excited-state methods like ADC(3) and CC3. Work is in progress towards that direction.

## Supporting Information

The size of the virtual space with different truncation thresholds for pyrrole, SS-FNO-EOM-CCSD excitation energy for Thiel, Charge transfer, and Waterloo test set, and the details of the THC-based treatment of the four external integrals are provided in the supporting information.

## Acknowledgments

The authors acknowledge the support from the IIT Bombay, CRG (Project no. CRG/2018/001549), and Matrix project of DST-SERB (Project No. MTR/2021/000420), CSIR-India (Project No. 01(3035)/21/EMR-II), UGC-India, DST-Inspire Faculty Fellowship (Project no. DST/INSPIRE/04/2017/001730), ISRO (Project No. RD/0122-ISROC00-004) for financial support. IIT Bombay super computational facility, and C-DAC Supercomputing resources (Param Smriti, Param Brahma) for computational time. AM acknowledges CSIR-HRDG for the research fellowship. AKD acknowledges the research fellowship funded by the EU NextGenerationEU through the Recovery and Resilience Plan for Slovakia under project No. 09I03-03-V04-00117.

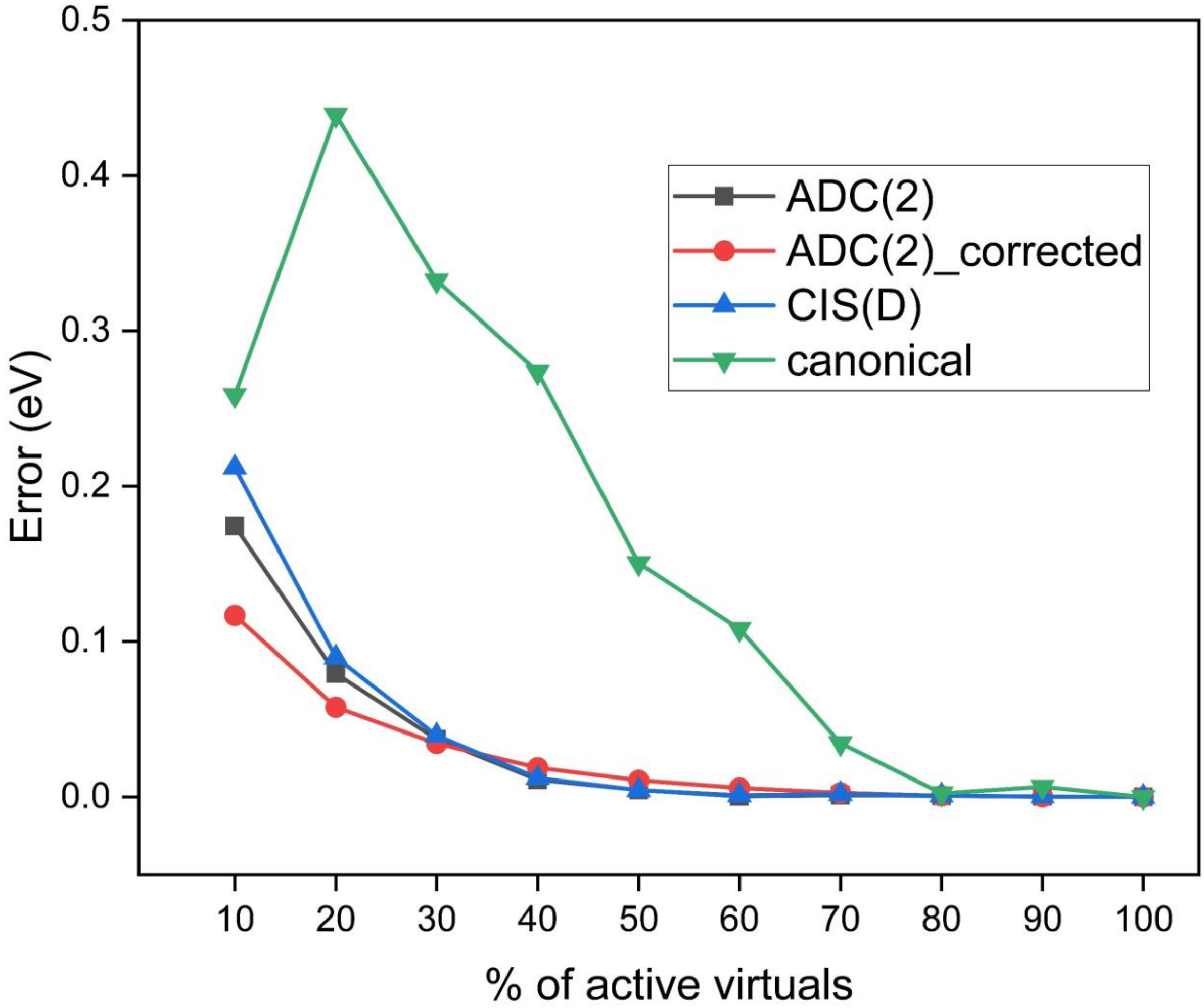

***Figure 1: The convergence error in SS-FNO-EE-EOM-CCSD and canonical EE-EOM-CCSD method with respect to percentage of active virtual $1^1B_2$ state of pyrrole molecule in aug-cc-pVTZ basis set. The canonical EE-EOM-CCSD result with 100% active virtual has been used as the reference.***

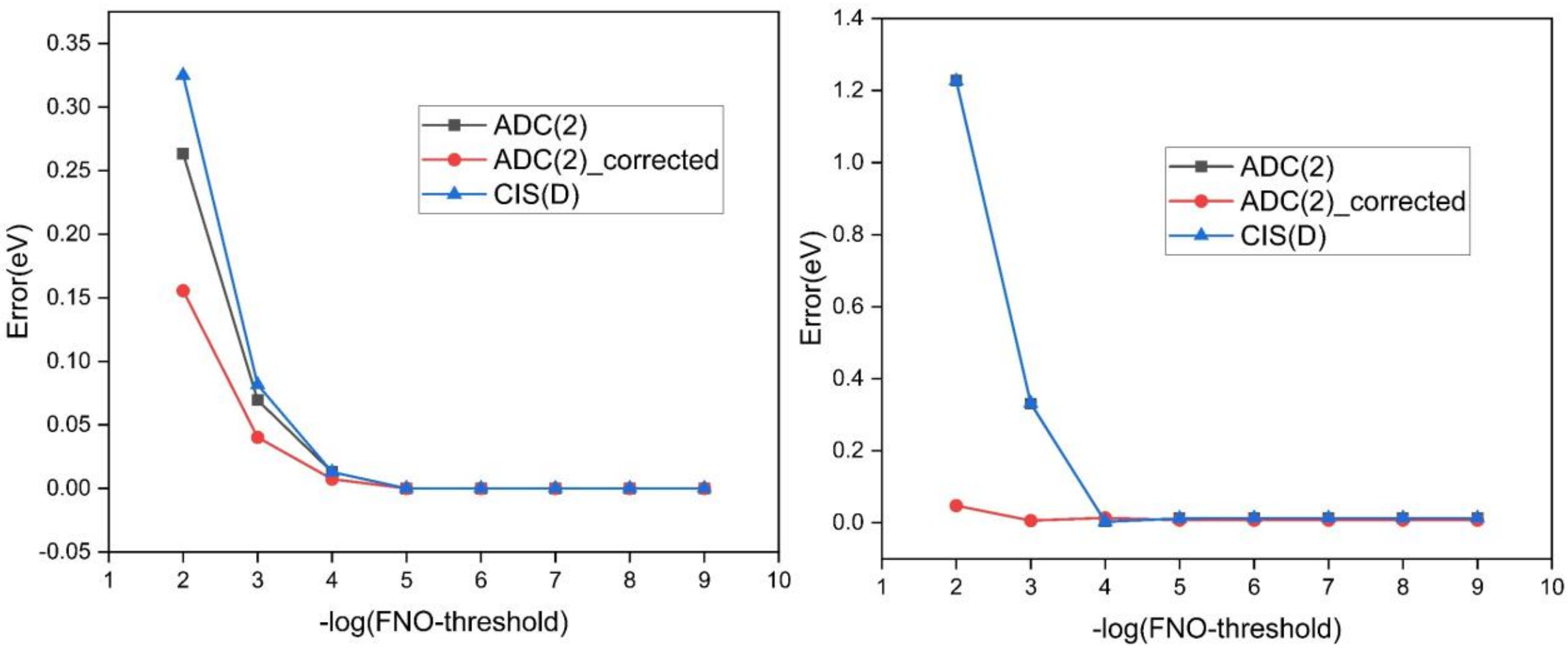

(a) (b)

***Figure 2: The convergence of error for the SS-FNO-EE-EOM-CCSD method for 1 $^{1}B^{2}$ valence excited state of pyrrole with respect to the (a) excited state FNO threshold keeping the ground state FNO threshold at zero(b) ground state FNO threshold, keeping the excited state FNO threshold at $10^{-4}$. The aug-cc-pVTZ basis has been used and the canonical EE-EOM-CCSD result with 100% active virtual has been used as the reference.***

**Table 1: Statistical Analysis of SS-FNO-EE-EOM-CCSD Excitation Energies Relative to Canonical EE-EOM-CCSD Values for Valence Excited State in Schreiber-Thiel's Test Set**

| | SS-FNO-EE-EOM-CCSD CIS(D) density | SS-FNO-EE-EOM-CCSD ADC (2) density | SS-FNO-EE-EOM-CCSD(Corrected) ADC (2) density |
|---|---|---|---|
| **ME** | 0.004 | 0.000 | 0.000 |
| **STD** | 0.008 | 0.006 | 0.004 |
| **RMSD** | 0.009 | 0.006 | 0.004 |
| **MAE** | 0.006 | 0.003 | 0.002 |
| **MAD** | 0.045 | 0.030 | 0.020 |

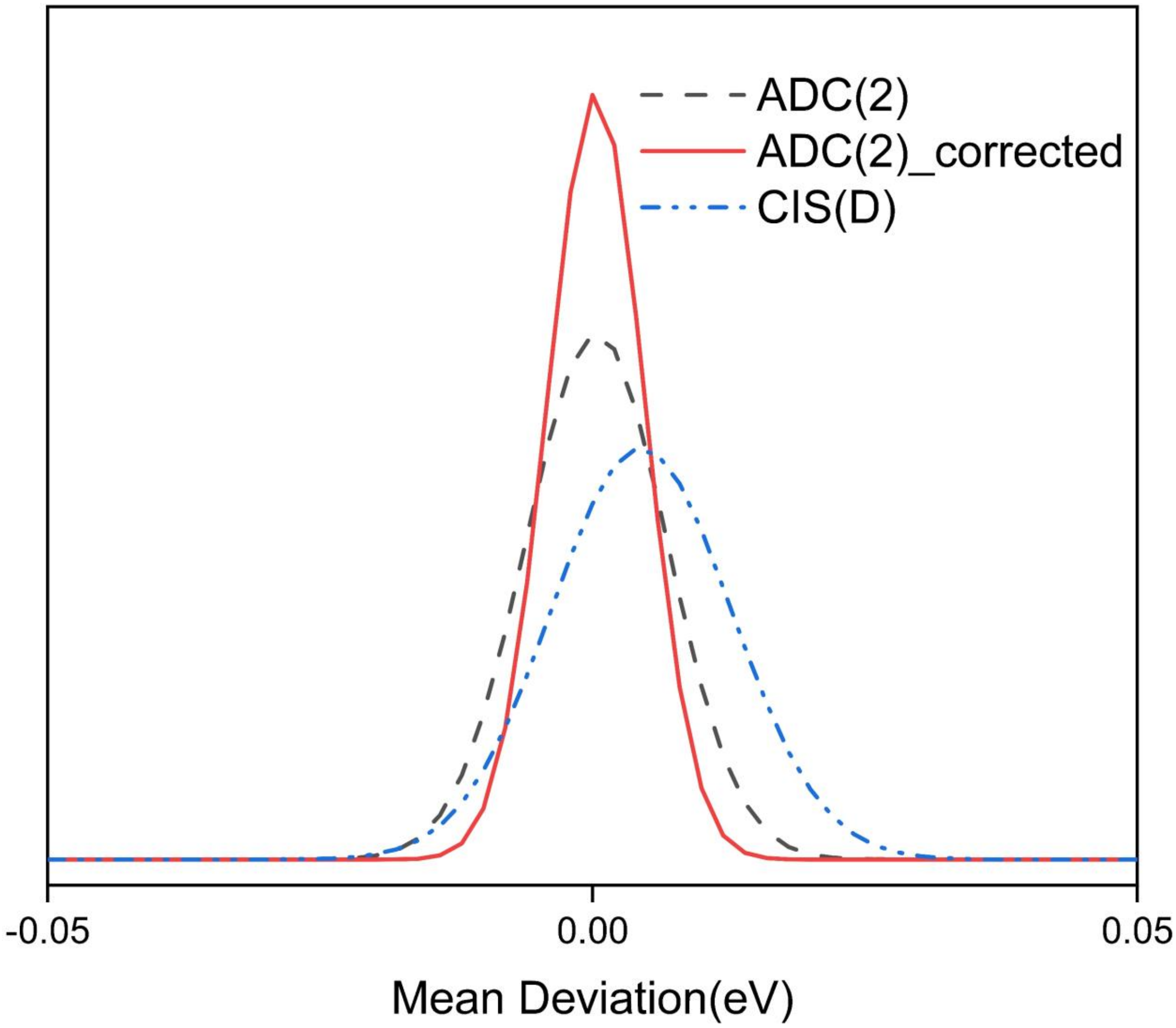

***Figure 3:*** *Error distributions in SS-FNO-EE-EOM-CCSD uncorrected and corrected results for valence excited states in Theil test set using CIS(D) and ADC(2) density. Errors are compared to canonical EE-EOM-CCSD, using TZVP basis set.*

**Table2: Statistical Analysis of SS-FNO-EE-EOM-CCSD Excitation Energies Relative to canonical EE-EOM-CCSD Values for Charge-transfer Test Set**

| | SS-FNO-EE-EOM-CCSD CIS(D) density | SS-FNO-EE-EOM-CCSD ADC (2) density | SS-FNO-EE-EOM-CCSD(Corrected) ADC (2) density |
|---|---|---|---|
| **ME** | -0.009 | -0.010 | -0.001 |
| **STD** | 0.024 | 0.025 | 0.013 |
| **RMSD** | 0.026 | 0.026 | 0.012 |
| **MAE** | 0.016 | 0.018 | 0.008 |
| **MAD** | 0.083 | 0.083 | 0.038 |

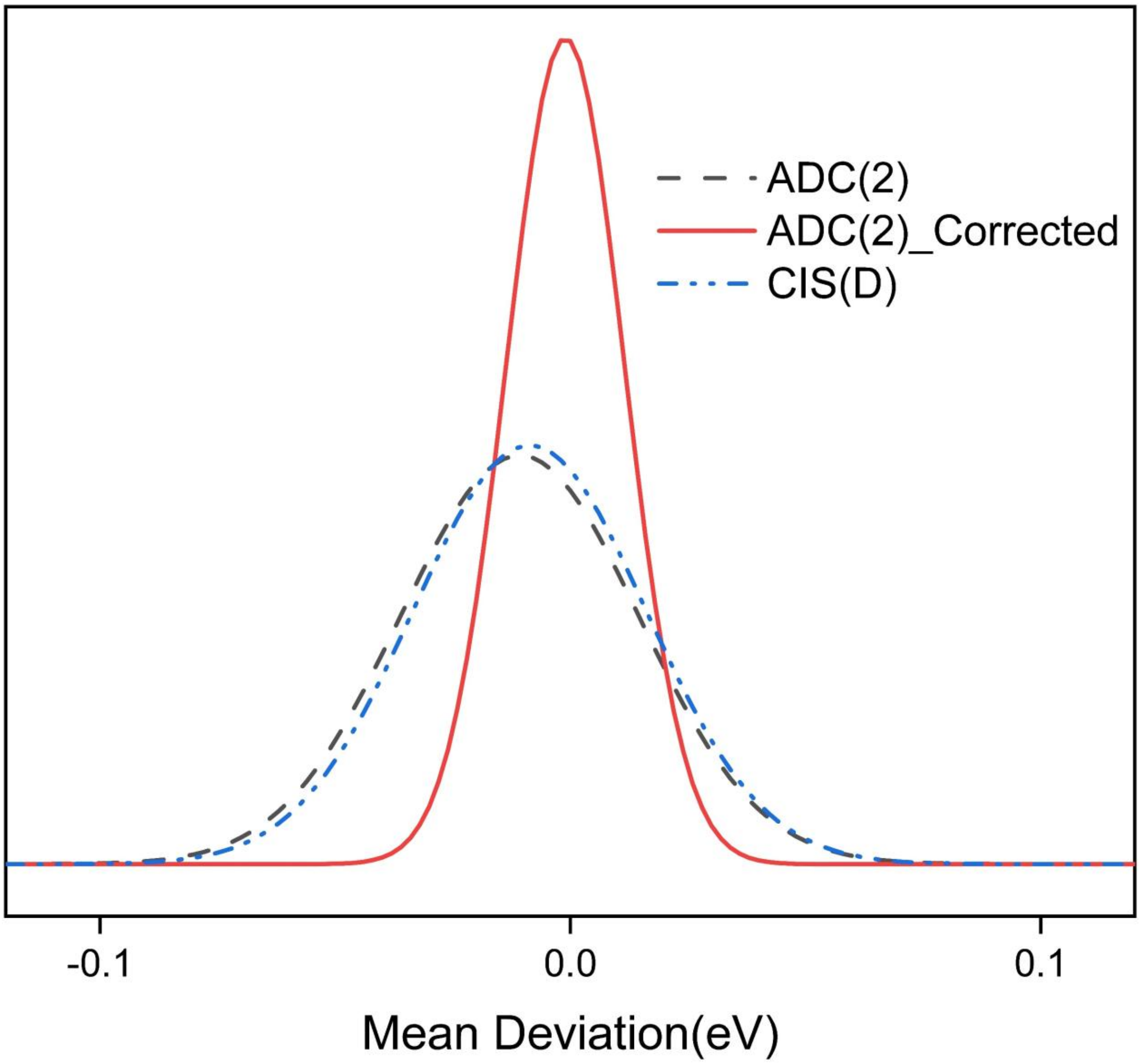

***Figure 4:*** *Error distributions in SS-FNO-EE-EOM-CCSD uncorrected and corrected results for the charge transfer test set using CIS(D) and ADC(2) density. Errors are compared to canonical EE-EOM-CCSD, using cc-pVDZ basis set.*

**Table 3: Statistical Analysis of SS-FNO-EE-EOM-CCSD Excitation Energies Relative to canonical EE-EOM-CCSD Values for Waterloo Rydberg Test Set:**

| | SS-FNO-EE-EOM-CCSD CIS(D) density | SS-FNO-EE-EOM-CCSD ADC (2) density | SS-FNO-EE-EOM-CCSD(Corrected) ADC (2) density |
|---|---|---|---|
| **ME** | 0.008 | 0.008 | 0.006 |
| **STD** | 0.019 | 0.015 | 0.014 |
| **RMSD** | 0.021 | 0.017 | 0.015 |
| **MAE** | 0.014 | 0.013 | 0.011 |
| **MAD** | 0.121 | 0.069 | 0.077 |

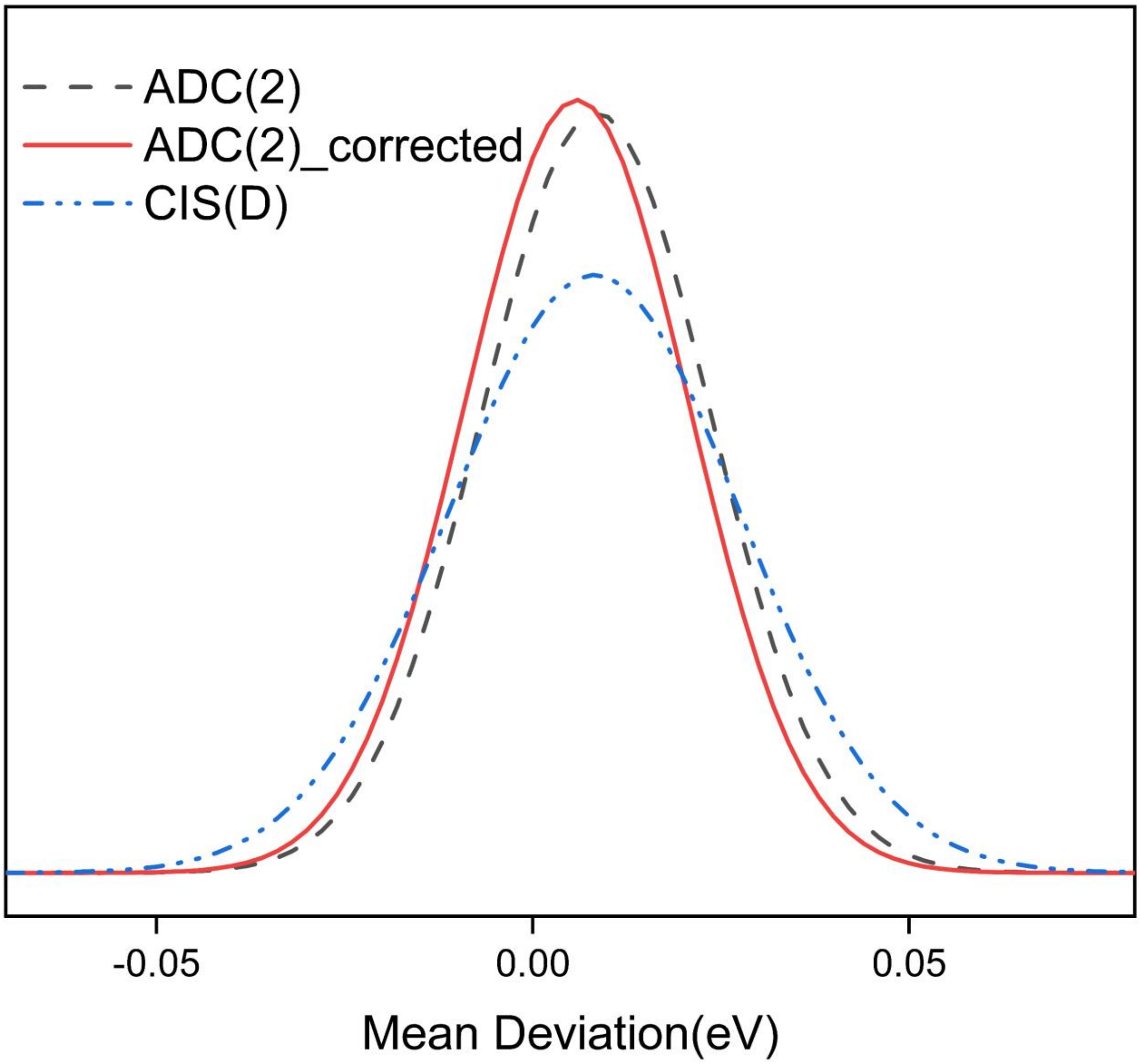

***Figure 5:*** *Error distributions in SS-FNO-EE-EOM-CCSD uncorrected and corrected results for Rydberg test set using CIS(D) and ADC(2) density. Errors are compared to canonical EE-EOM-CCSD, using TZVP (with additional 5s, 4p, and 4d functions added to center of mass) basis set.*

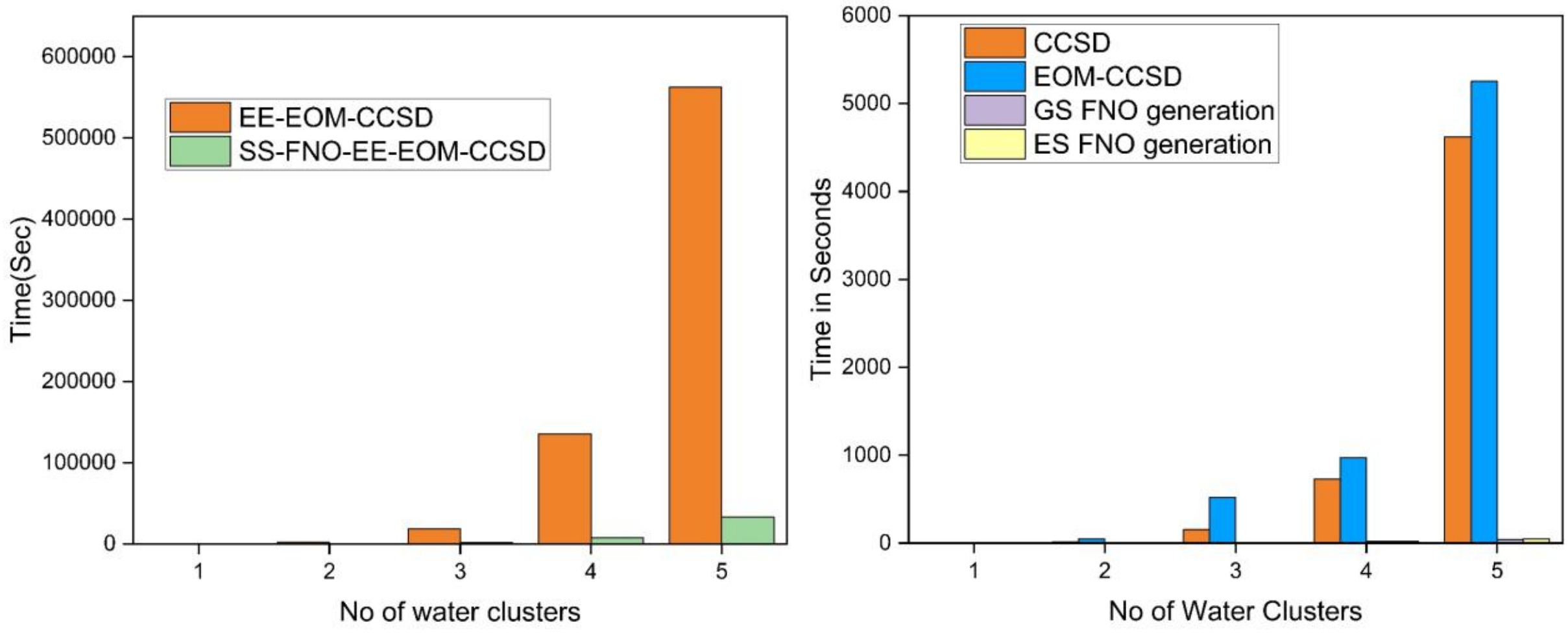

(a) (b)

*Figure 6: (a)comparison of time taken between canonical EE-EOM-CCSD and SS-FNO-EE-EOM-CCSD methods (b)the time taken by individual steps in SS-FNO-EOM-CCSD, for water clusters ranging from monomer to pentamer (*$(H_2O)_n$*, n= 1,2,..,5)*

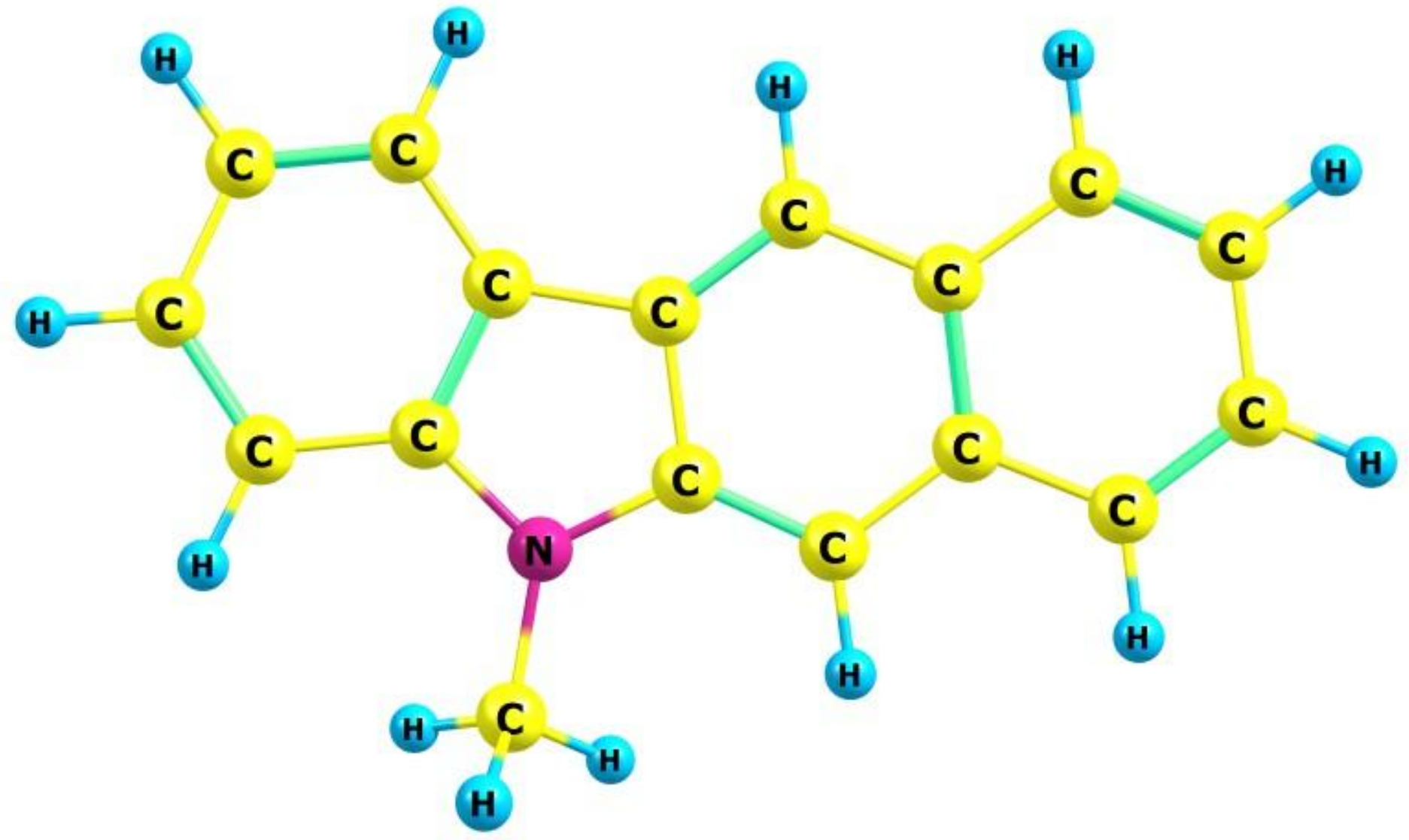

*Figure 7: Structure of N-methyl-2,3-benzcarbaxole.*